\documentclass{ws-mpla}
\newcommand{\calL}{\mathcal{L}}
\begin{document}

\markboth{T. J. Marshall, D. G. C. McKeon} {Radiative Properties
of the Stueckelberg Mechanism}

\title{Radiative Properties of the Stueckelberg Mechanism}

\author{\footnotesize  T. J. MARSHALL}
\address{Department of Applied Mathematics, University of Western Ontario\\
Middlesex College Room 210, London, N6H 1B7, Canada}

\author{\footnotesize D. G. C. MCKEON}
\address{Department of Applied Mathematics, University of Western Ontario\\
Middlesex College Room 269, London, N6H 1B7, Canada\\
dgmckeo2@uwo.ca}

\maketitle

\begin{abstract}
We examine the mechanism for generating a mass for a U(1) vector
field introduced by Stueckelberg.  First, it is shown that
renormalization of the vector mass is identical to the
renormalization of the vector field on account of gauge
invariance.  We then consider how the vector mass affects the
effective potential in scalar quantum electrodynamics at one-loop
order.  The possibility of extending this mechanism to couple, in
a gauge invariant way, a charged vector field to the photon is
discussed.
\end{abstract}


\ccode{PACS Nos.: }

\section{Introduction}

\normalsize The masslessness of the photon is well established;
indeed experiment shows that $m_\gamma < 3\times10^{-27}$eV.
However, the vector Bosons associated with the weak interactions
must be massive.  It has been shown how the non-Abelian gauge
Bosons associated with the electroweak interactions can be given
this mass in a way that retains gauge invariance through the Higgs
mechanism, thereby ensuring the renormalizability of the model\cite{t'Hooft}.\\
\indent There is, however, a way of providing a mass to a $U(1)$
vector Boson that retains renormalizability and unitarity without
the use of the Higgs mechanism.  This ``Stueckelberg
mechanism"\cite{Stueckelberg} does not involve the presence of an
extra degree of freedom in the physical spectrum, in contrast to the Higgs mechanism.\\
\indent In Ref. 3 and 4, the possibility of the vector Boson
associated with the $U(1)$ sector of the Standard Model acquiring
a mass through the Stueckelberg mechanism was examined.  There it
was demonstrated that the mass matrix associated with the vector
Boson in the full $SU(2)\times U(1)$ electroweak model does not
have a vanishing eigenvalue if there is only a single $U(1)$
vector and a single Higgs doublet and consequently there is no
massless photon. This means that if the Stueckelberg mass is
non-zero in addition to the usual Standard Model parameters, the
photon cannot be massless. There is no reason why the possibility
of having a non-vanishing photon mass should be discarded within
the content of the usual Standard Model; a non-vanishing
Stueckelberg mass is consistent with renormalizability and
unitarity. When dealing with the usual Standard Model, there is no
more reason for setting the photon mass equal to zero than there
is for setting the cosmological constant equal to zero.  This
forces one to consider extensions of the Standard Model to
reconcile it with the observed masslessness of the photon, if
indeed ``what is not forbidden must be
allowed."\\
\indent The Stueckelberg mechanism can be used in conjunction with
other models involving gauge symmetries.  Indeed, its extensions
to supersymmetric gauge models\cite{Kuzmin and McKeon 2,Kors and
Nath 2} and to spin-two models\cite{Delbourge and Salam,Dilkes}
have been considered. It
also arises naturally in effective actions generated by string models\cite{Kors and Nath 3}.\\
\indent In Ref. 4 and 9, the Standard Model is extended so as to
accommodate a massless photon when the Stueckelberg mechanism
occurs by including a second $U(1)$ sector.  This results in the
mass matrix for the vectors having a vanishing eigenvalue, so that
a massless photon can be incorporated into the model even when a
Stueckelberg mass arises.  Another extension of the Standard Model
that ensures that a massless photon occurs is to embed the $U(1)$
symmetry of the Standard Model into a larger non-Abelian gauge
group such as in the Grand Unified $SU(5)$ model.  Since one
cannot generalize the Stueckelberg mechanism so as to accommodate
a non-Abelian gauge symmetry, the $U(1)$ sector of the Standard
Model cannot be associated with the Stueckelberg mechanism when
this $U(1)$ gauge symmetry is just a remnant of the larger
non-Abelian symmetry once its symmetry is broken.\\
\indent With the Stueckelberg mechanism having possible
application in the Standard Model, it is relevant to consider some
of its field theoretical consequences.  We first examine how the
Stueckelberg mass is renormalized, showing that on account of
gauge invariance, this mass renormalization is dictated by the
renormalization of the photon wave function and consequently of
the $U(1)$ gauge coupling constant.  Unlike other masses in the
Standard Model it is not renormalized independently of other
quantities that occur.\\
\indent Next we demonstrate how in scalar electrodynamics, the
presence of a Stueckelberg mass for the vector field considerably
alters the form of the radiatively generated effective potential
in the model.\\
\indent Finally, it is shown how the Stueckelberg field can be
used not only as to permit one to introduce a mass for a $U(1)$
vector field, but also to couple a complex vector field to a
photon in a way that preserves gauge invariance for the complex
vector field.  Unfortunately, the resulting model, through gauge
invariant, is not renormalizable.  It thus appears that only
through the Higgs mechanism can a massive charged vector field
arise if one is only accepting of renormalizable models.\\

\section{Renormalization of the Stueckelberg Mass}

The usual Maxwell Lagrangian can be supplemented by a gauge
invariant mass term to yield the Stueckelberg Lagrangian
\begin{equation}
\calL_s = -\frac{1}{4}\left(\partial_\mu A_\nu-\partial_\nu
A_\mu\right)^2 +
\frac{1}{2}m_s^2\left(A_\mu+\frac{1}{m_s}\partial_\mu\sigma\right)^2;
\end{equation}
this possess the gauge invariance
\begin{eqnarray}
\nonumber A_\mu &\to& A_\mu+\partial_\mu\Omega\\
\sigma&\to&\sigma-m_s\Omega.
\end{eqnarray}
If Eq.~(1) is supplemented by the gauge fixing term,
\begin{equation}
\calL_{gf}=-\frac{1}{2\xi}(\partial\cdot A-\xi m_s\sigma)^2
\end{equation}
then $A_\mu$ and $\sigma$ decouple in $\calL_s+\calL_{gf}$.\\
\indent The propagator for the field $A_\mu$ is
\begin{equation}
\left <A_\mu A_\nu\right > =
\frac{-i}{k^2-m_s^2}\left(g_{\mu\nu}-\frac{(1-\xi)k_\mu
k_\nu}{k^2-\xi m_s^2}\right).
\end{equation}
Renormalizability is manifest if $\xi=1$, when $\xi\to\infty$ we
recover the usual propagator for a massive vector, and when
$\xi=0$ we are in a ``unitary" gauge in which only the
transverse degrees of freedom of $A_\mu$ contribute.\\
\indent If we now couple $A_\mu$ to a spinor field $\psi$ so that
we have in addition to $\calL_s$ and $\calL_{gf}$
\begin{equation}
\calL_\psi=\stackrel{\_}{\psi}[(i\partial_\mu-eA_\mu)\gamma^\mu-m]\psi
\end{equation}
then the gauge transformations (2) are accompanied by
\begin{equation}
\psi\to e^{-ie\Omega}\psi.
\end{equation}
On account of the gauge invariance of $\calL_s+\calL_\psi$, the
usual Ward-Takahashi-Slavnov-Taylor (WTST) identities of quantum
electrodynamics (QED) persist even when $m_s^2\not=0$.  As a
result, the regulated one particle irreducible two point
function$\left<A_\mu A_\nu\right>$ in momentum space,
$\pi_{\mu\nu}(k)$, is of the form
\begin{equation}
i\pi_{\mu\nu}(k)=i(g_{\mu\nu}k^2-k_\mu k_\nu)\pi(k^2)\equiv
ig^T_{\mu\nu}k^2\pi(k^2).
\end{equation}
Working in the gauge $\xi=0$, iteration of this contribution to
the two point function leads to
\begin{eqnarray*}
\frac{-ig^T_{\mu\nu}}{k^2-m_s^2}+
\frac{-ig^T_{\mu\lambda}}{k^2-m_s^2}(i\pi_{\lambda\sigma})
\frac{-ig^T_{\sigma\nu}}{k^2-m_s^2}
+\frac{-ig^T_{\mu\lambda}}{k^2-m_s^2}(i\pi_{\lambda\sigma})
\frac{-ig^T_{\sigma\rho}}{k^2-m_s^2}(i\pi_{\rho\kappa})\frac{-ig^T_{\kappa\nu}}{k^2-m_s^2}
+\ldots
\end{eqnarray*}
\begin{eqnarray}
=\frac{-ig_{\mu\nu}^T\left(\frac{1}{1-\pi}\right)}{k^2-m_s^2\left(\frac{1}{1-\pi}\right)}.
\end{eqnarray}
From Eq.~(8) we see that divergences that appear in $\pi(k^2)$
when the regulating parameter approaches its limiting value appear
in two places; those in the numerator of Eq.~(8) serve to
renormalize the external wave function, while those in the
denominator renormalize $m_s^2$.  The WTST identity that relates
the vertex wave function to the spinor self energy, relates this
renormalization of the wave function to the renormalization of the
coupling constant $e^2$.  In fact, renormalization of the electric
charge is given by,
\begin{equation}
e^2_R=e^2(1+\pi)^{-1}_{div}
\end{equation}
and also by Eq.~(8),
\begin{equation}
(m_s^2)_R=m_s^2(1+\pi)^{-1}_{div}
\end{equation}
where $(1+\pi)^{-1}_{div}$ indicates the divergent contribution to
$(1+\pi)^{-1}$ that arises when the regulating parameter
approaches its limiting value.  The finite renormalized coupling
and mass are $e^2_R$ and $(m_s^2)_R$ respectively.  It follows
from Eqs.~(8) and (9) that the renormalization group functions
that dictate how $e_R^2$ and $(m_s^2)_R$ vary with the
renormalization scale are identical.

\section{The Effective Action}

The one-loop radiative corrections to the effective action in
scalar electrodynamics have been considered in Ref. 10.  We here
consider how the inclusion of a Stueckelberg mass into the model
affects this calculation.\\
\indent The Lagrangian $\calL_s$ of (1) is supplemented with a
Lagrangian which couples $A_\mu$ to a complex scalar field
$\phi(x)$,
\begin{equation}
\calL_\phi=(\partial_\mu+ieA_\mu)\phi^*(\partial^\mu-ieA^\mu)\phi-\kappa^2\phi^*\phi-\lambda(\phi^*\phi)^2
\end{equation}
If we assume that $\phi(x)$ has a constant background component
$f$, taken to be real, then
\begin{equation}
\sqrt{2}\phi(x)=f+h_1(x)+ih_2(x)
\end{equation}
and we find that the most convenient gauge fixing Lagrangian is no
longer Eq.~(3) but rather,
\begin{equation}
\calL_{\stackrel{\_}{gf}}=-\frac{1}{2\xi}[\partial\cdot
A-\xi(m_s\sigma-efh_2)]^2.
\end{equation}
The terms in $\calL_s+\calL_\phi+\calL_{\stackrel{\_}{gf}}$ that
are bilinear in the quantum fields in the gauge where $\xi=1$ are,
\begin{eqnarray}
\calL^{(2)}=\frac{1}{2}(h_1,h_2,\sigma,A_\mu)\bf H\rm
\left(\begin{array}{c}h_2\\h_1\\\sigma\\A_\nu\end{array}\right)
\end{eqnarray}
where
\begin{eqnarray*}
\bf H\rm=\left(\begin{array}{cccc}
p^2-(\kappa^2+3\lambda f^2)&0&0&0\\
0&p^2-(\kappa^2+\lambda f^2+e^2f^2)&\frac{1}{2}m_sef&0\\
0&\frac{1}{2}m_sef&p^2-m_s^2&0\\
0&0&0&g^{\mu\nu}(-p^2+m_s^2+e^2f^2)\end{array}\right)
\end{eqnarray*}
The one-loop effective potential is now given by\cite{Jackiw},
\begin{equation}
V^{(1)}=-\ln(\det \bf H\rm)^{-\frac{1}{2}},
\end{equation}
where $\bf H\rm$ is the functional matrix appearing in Eq.~(14).
We note that on account of the form of the gauge fixing term in
Eq.~(13) the Stueckelberg field $\sigma$ does not decouple from
the other fields.  This gauge fixing does, however, ensure
that $A_\mu$ decouples in $\bf H\rm$.\\
\indent Diagonalizing the matrix $\bf H\rm$, we find that,
\begin{equation}
V^{(1)}=-\ln\left[\det\left(\begin{array}{cccc}
p^2-(\kappa^2+3\lambda f^2)&0&0&0\\
0&g_{\mu\nu}(p^2-m_s^2-e^2f^2)&0&0\\
0&0&p^2-m_+^2&0\\
0&0&0&p^2-m_-^2\end{array}\right)\right]^{-\frac{1}{2}}
\end{equation}
where,
\begin{equation}
2m_\pm^2=[m_s^2+\kappa^2+(\lambda+e^2)f^2]\pm\sqrt{[\kappa^2+(\lambda+e^2)f^2-m_s^2]^2+e^2m_s^2f^2}.
\end{equation}
The functional determinant in Eq.~(16) can be evaluated using
operator regularization, a variant of $\zeta$-function
regularization\cite{McKeon and Sherry 1,McKeon and Sherry 2},
which preserves symmetries and circumvents all explicit
divergences. With this we find that
\begin{equation}
V^{(1)}=-\frac{1}{2}\lim_{s\to
0}\frac{d}{ds}\frac{\mu^{2s}}{\Gamma(s)}tr\int^{\infty}_0
d(it)(it)^{(s-1)}e^{(it\bf H\rm^D)}
\end{equation}
where $\bf H\rm^D$ is the diagonal matrix in Eq.~(16).\\
\indent Since
\begin{equation}
tr\left(e^{ip^2t}\right)=\int\frac{d^4p}{(2\pi)^4}e^{ip^2t}=\frac{i}{(4\pi
it)^2}
\end{equation}
and
\begin{equation}
\int_0^{\infty}d(it)(it)^{s-3}e^{-im^2t}=\Gamma(s-2)\left(m^2\right)^{2-s}
\end{equation}
Eq.~(18) reduces to,
\begin{eqnarray*}
\nonumber V^{(1)}=-\frac{1}{32\pi^2}\lim_{s\to
0}\frac{d}{ds}\frac{\mu^{2s}\Gamma(s-2)}{\Gamma(s)}\left[(\kappa^2+3\lambda
f^2)^{2-s}+4(m_s^2+e^2f^2)^{2-s}+\left(m_+^2\right)^{2-s}+\left(m_-^2\right)^{2-s}\right]
\end{eqnarray*}
\begin{eqnarray*} \nonumber=\frac{1}{64\pi^2}\left[(\kappa^2+3\lambda
f^3)^2\left(\ln\left(\frac{\kappa^2+3\lambda
f^2}{\mu^2}\right)-\frac{3}{2}\right)+
4(m_s^2+e^2f^2)^2\left(\ln\left(\frac{m_s^2+e^2f^2}{\mu^2}\right)-\frac{3}{2}\right)\right.
\end{eqnarray*}
\begin{eqnarray}
\left.+m_+^2\left(\ln\left(\frac{m_+^2}{\mu^2}\right)-\frac{3}{2}\right)+
m_-^2\left(\ln\left(\frac{m_-^2}{\mu^2}\right)-\frac{3}{2}\right)\right].
\end{eqnarray}
Supplementing $V^{(1)}$ with $V^{(0)}=\frac{\lambda f^2}{4}$ to
form the effective potential $V(f)=V^{(0)}+V^{(1)}$ leads to
rather complicated dependence of $V$ on $f$, especially when
$m_s^2\not=0$.  Minimizing $V$ at $f=v$ leads to a vacuum
expectation value of $\phi(x)$.

\section{Charged Vector Field}

A complex vector field $W\mu$ with action,
\begin{eqnarray*}
\calL_W=-\frac{1}{2}(\partial_\mu W_\nu^*-\partial_\nu
W_\mu^*)(\partial^\mu W^\nu-\partial^\nu W^\mu)
\end{eqnarray*}
\begin{eqnarray}
+m_W^2\left(W_\mu^*+\frac{1}{m_W}\partial_\mu\Sigma^*\right)
\left(W^\mu+\frac{1}{m_W}\partial^\mu\Sigma\right),
\end{eqnarray}
possesses the gauge invariance
\begin{eqnarray}
\nonumber W_\mu&\to&W_\mu+\partial_\mu\omega\\
\Sigma&\to&\Sigma-m_W\omega
\end{eqnarray}
where $\Sigma$ is a complex scalar and $\omega$ is a complex gauge
function.  Coupling this vector field to a massive photon through
replacement of the ordinary derivative by a covariant derivative
leads to\cite{Kuzmin and McKeon 3},
\begin{eqnarray*}
\calL_{WA}=-\frac{1}{4}(\partial_\mu A_\nu-\partial_\nu A_\mu)^2
\end{eqnarray*}
\begin{eqnarray*}
-\frac{1}{2}\left[(\partial_\mu+ieA_\mu)W_\nu^*-(\partial_\nu+ieA_\nu)W_\mu^*\right]
\left[(\partial^\mu-ieA^\mu)W^\nu-(\partial^\nu-ieA^\nu)W^\mu\right]
\end{eqnarray*}
\begin{eqnarray}
+\frac{m_A^2}{2}\left(A_\mu+\frac{1}{m_A}\partial_\mu\sigma\right)^2
+m_W^2\left(W_\mu^*+\frac{1}{m_W}\partial_\mu\Sigma^*\right)
\left(W^{\mu}+\frac{1}{m_W}\partial^\mu\Sigma\right)
\end{eqnarray}
This is invariant the gauge transformations
\begin{eqnarray}
\nonumber\sigma&\to&\sigma-m_A\theta\\
W_\mu&\to&e^{ie\theta}W_\mu\\
\nonumber A_\mu&\to&A_\mu+\partial_\mu\theta
\end{eqnarray}
Regrettably the gauge transformation of Eq.~(23) is broken.
However, the Lagrangian,
\begin{eqnarray*}
\calL=-\frac{1}{4}(\partial_\mu A_\nu-\partial_\nu
A_\mu)^2-\frac{1}{2}\left[(\partial_\mu+ieA_\mu)W_\nu^*-
(\partial_\nu+ieA_\nu)W_\mu^*+\frac{ie}{m_W}(\partial_\mu
A_\nu-\partial_\nu A_\mu)\Sigma^*\right]
\end{eqnarray*}
\begin{eqnarray*}
\left[(\partial^\mu-ieA^\mu)W^\nu-
(\partial^\nu-ieA^\nu)W^\mu-\frac{ie}{m_W}(\partial^\mu
A^\nu-\partial^\nu A^\mu)\Sigma\right]
\end{eqnarray*}
\begin{eqnarray*}
+m_W^2\left[W_\mu^*+\frac{1}{m_W}(\partial_\mu+ieA_\mu)\Sigma^*\right]
\left[W^\mu+\frac{1}{m_W}(\partial^\mu-ieA^\mu)\Sigma\right]
\end{eqnarray*}
\begin{eqnarray}
+\frac{m_A^2}{2}(A_\mu+\frac{1}{m_A}\partial_\mu\sigma)^2
\end{eqnarray}
does in fact possess the gauge invariance
\begin{eqnarray}
\nonumber W_\mu&\to&W_\mu+(\partial_\mu-ieA_\mu)\omega\\
A_\mu&\to&A_\mu\\
\nonumber\Sigma&\to&\Sigma-m_W\omega
\end{eqnarray}
as well as that of Eq.~(25).\\
\indent An obvious generalization of the gauge fixing Lagrangian
(3) is,
\begin{eqnarray*}
\calL_{gf}=-\frac{1}{2\xi}(\partial\cdot A-\xi m_A\sigma)^2
\end{eqnarray*}
\begin{eqnarray}
-\frac{1}{\zeta}\left[(\partial+ieA)\cdot W^*-\zeta
m_W\Sigma^*\right] \left[(\partial-ieA)\cdot W-\zeta
m_W\Sigma\right].
\end{eqnarray}
Unfortunately, this gauge fixing Lagrangian does not serve to
completely decouple the Stueckelberg field $\Sigma$.  There still
remains the coupling
\begin{equation}
\calL_I=-\frac{e^2}{2m_W^2}(\partial_\mu A_\nu-\partial_\nu
A_\mu)^2\Sigma^*\Sigma
\end{equation}
which destroys renormalizability.  This results in a counter term
proportional to\\$\left[(\partial_\mu A_\nu-\partial_\nu
A_\mu)(\partial^\mu A^\nu-\partial^\nu A^\mu)\right]^2$ being
required, even at one loop order.  It appears that in order to
couple a photon $A_\mu$ to a massive vector field $W_\mu$, one
requires an $O(3)$ Yang Mills interaction, with the Higgs
mechanism used to provide a mass to the field $W_\mu$, if one is
to retain renormalizability (two of the components of the $O(3)$
gauge field are used to compose $W_\mu$ and $W^*_\mu$; the third
component is identified with $A_\mu$).

\section{Conclusion}

We have examined several aspects of the Stueckelberg mechanism for
generating a mass for a $U(1)$ vector field.  First of all, we
have demonstrated that renormalization of the vector field is
proportional to that of the Stueckelberg mass.  Next we have shown
how the presence of a Stueckelberg mass affects the one-loop
effective potential in scalar electrodynamics.  Finally we have
attempted (unsuccessfully) to use an extension of the Stueckelberg
mechanism to formulate a renormalizable model for a charged
massive vector field.

\section*{Acknowledgements}

NSERC for providing financial support.  Roger Macloud for his
helpful suggestion.

\end{document}